\documentclass[
reprint,           % journal's actual layout
twocolumn,
superscriptaddress,% authors with affiliations via superscripts
amsmath,           % add AMS-Latex features
amssymb,           % add extra AMS symbols, including amsfonts
aps,               % aps or aip
prd,               % prl, pra, prb, prc, prd, pre, prstab
notitlepage,       % control appearance of title page
longbibliography,  % show  article titles in the bibliography
floatfix,          % process floats as early as possible
nofootinbib,
]{revtex4-1}

\usepackage{amsmath}
\allowdisplaybreaks[4]
\usepackage{cancel}
\usepackage{extarrows}
\usepackage{tensor}     % manipulate tensors
\usepackage{float}
\usepackage[caption = false]{subfig} % combine multiple figures
\usepackage[final]{graphicx}   % include figures
\usepackage[
colorlinks=true,        % color link
citecolor=blue,         % cite color
linkcolor=blue,         % link color
urlcolor=blue           % url color
]{hyperref}             % create hyperlinks
\usepackage{bm}         % \bm{<text>} Bold math symbols
\usepackage{xcolor}     % textcolor
\usepackage{lipsum}
\usepackage{color}      % \textcolor{declared-color}{text}
\usepackage[utf8]{inputenc} % accented characters for .bib file using XeLaTex
\usepackage[section]{placeins} % figures processing
\usepackage{appendix}
\usepackage{units}
\newcommand{\nc}{\newcommand*} 
\newcommand{\apjl}{Ap.J.L.}
\newcommand{\pasa}{Publ. Astro. Soc. Aust.}

\nc{\figurewidth}{3.2in}
\nc{\xbar}{\bar{x}}
\nc{\rhoeq}{\rho_{\mathrm{eq}}}
\nc{\zeq}{z_{\mathrm{eq}}}
\nc{\tla}{\tilde{\lambda}}
\nc{\dt}{\delta}
\nc{\Dt}{\Delta}
\nc{\vj}{\hat{j}}
\nc{\vl}{\hat{l}}
\nc{\hx}{\hat{x}}
\nc{\hy}{\hat{y}}
\nc{\bj}{\bm{j}}
\nc{\mJ}{\mathcal{J}}
\nc{\mP}{\mathcal{P}}
\nc{\Msun}{M_\odot}
\nc{\app}{\approx}
\nc{\av}[1]{\langle #1 \rangle}
\nc{\eq}[1]{Eq.~\eqref{#1}}
\nc{\al}{\alpha}
\nc{\Xstar}{X_{\ast}}
\nc{\seq}{\sigma_{\mathrm{eq}}}
\nc{\fpbh}{f_{\mathrm{pbh}}}
\nc{\vth}{\hat{\theta}}
\nc{\vla}{\hat{\lambda}}
\nc{\vd}{\hat{d}}
\nc{\Mmin}{M_{\mathrm{min}}}
\nc{\rmd}{\mathrm{d}}
\nc{\mmin}{{m_{\mathrm{min}}}}
\nc{\mmax}{{m_{\mathrm{max}}}}
\nc{\mR}{\mathcal{R}}
\nc{\tmR}{\tilde{\mathcal{R}}}
\nc{\s}{\sigma}
\nc{\ogw}{\Omega_{\mathrm{GW}}}
\nc{\addref}{[\textcolor{red}{add ref}] }
\nc{\Om}{\Omega}
\nc{\gm}{\gamma}
\nc{\Gm}{\Gamma}
\nc{\gpcyr}{\mathrm{Gpc}^{-3}\,\mathrm{yr}^{-1}}
\nc{\Eq}[1]{Eq.~\eqref{#1}}
\nc{\Fig}[1]{Fig.~\ref{#1}}
\nc{\Table}[1]{Table~\ref{#1}}
\nc{\lvc}{LIGO/Virgo} % LIGO-VIRGO collaboration
\nc{\Sec}[1]{Sec.~\ref{#1}}
\nc{\eg}{\textit{e.g.~}}
\nc{\SNR}{\mathrm{SNR}}
\nc{\bt}{\mathbf{t}}
\nc{\be}{\mathbf{\epsilon}}
\nc{\bn}{\mathbf{n}}
\nc{\bx}{\mathbf{x}}
\nc{\bk}{\mathbf{k}}
\nc{\bd}{\mathbf{d}}
\nc{\ba}{\mathbf{a}}
\nc{\bp}{\mathbf{p}}
\nc{\bnu}{\mathbf{\nu}}
\nc{\uni}{\mathrm{U}}
\nc{\logu}{\operatorname{\mathrm{log-U}}}
\nc{\RN}{\mathrm{RN}}
\nc{\BN}{\mathrm{BN}}
\nc{\GN}{\mathrm{GN}}
\nc{\mcN}{\mathcal{N}}
\nc{\GWB}{\mathrm{GW}}
\nc{\yr}{\mathrm{yr}}
\nc{\Am}{\mathcal{A}}
\nc{\Dm}{\mathcal{D}}
\nc{\Hm}{\mathcal{H}}
\nc{\sovast}{Soviet Ast.}
\nc{\hosc}{h_{\mathrm{osc}}}
\nc{\Posc}{\Psi_{\mathrm{osc}}}

\nc{\mrm}{\mathrm}
\nc{\BE}{B\scriptsize{AYES}\normalsize{E}\scriptsize{PHEM}\normalsize  }

\def\({\left(}
\def\){\right)}
\def\[{\left[}
\def\]{\right]}

\def\e{\begin{equation}}
\def\q{\end{equation}}
\def\m{\begin{eqnarray}}
\def\n{\end{eqnarray}}
\nc{\red}[1]{\textcolor{red}{#1}}
\begin{document}

\title{Constraining ultralight vector dark matter with the Parkes Pulsar Timing Array second data release}     

%%%%%%%%%%%%%%%%%%%%%%%%%%%%%%%%%%%% author %%%%%%%%%%%%%%%%%%%%%%%%%%%%%%%%%%%%
\author{Yu-Mei Wu}
\email{wuyumei@itp.ac.cn} 
\affiliation{School of Fundamental Physics and Mathematical Sciences, Hangzhou Institute for Advanced Study, UCAS, Hangzhou 310024, China}
\affiliation{School of Physical Sciences, University of Chinese Academy of Sciences, No. 19A Yuquan Road, Beijing 100049, China}
\affiliation{CAS Key Laboratory of Theoretical Physics, Institute of Theoretical Physics, Chinese Academy of Sciences, Beijing 100190, China}

%%%%%%%%%%%%%%%%%%%%%%%%%%%%%%%%%%%% author %%%%%%%%%%%%%%%%%%%%%%%%%%%%%%%%%%%%
\author{Zu-Cheng Chen}
\email{Corresponding author: zucheng.chen@bnu.edu.cn}
\affiliation{Department of Astronomy, Beijing Normal University, Beijing 100875, China}
\affiliation{Advanced Institute of Natural Sciences, Beijing Normal University, Zhuhai 519087, China}

%%%%%%%%%%%%%%%%%%%%%%%%%%%%%%%%%%%% author %%%%%%%%%%%%%%%%%%%%%%%%%%%%%%%%%%%%
\author{Qing-Guo Huang}
\email{Corresponding author: huangqg@itp.ac.cn}
\affiliation{School of Fundamental Physics and Mathematical Sciences, Hangzhou Institute for Advanced Study, UCAS, Hangzhou 310024, China}
\affiliation{CAS Key Laboratory of Theoretical Physics, 
    Institute of Theoretical Physics, Chinese Academy of Sciences,
    Beijing 100190, China}
\affiliation{School of Physical Sciences, 
    University of Chinese Academy of Sciences, 
    No. 19A Yuquan Road, Beijing 100049, China}

%%%%%%%%%%%%%%%%%%%%%%%%%%%%%%%%%%%%% author %%%%%%%%%%%%%%%%%%%%%%%%%%%%%%%%%%%%
\author{Xingjiang Zhu}
\email{Corresponding author: zhuxj@bnu.edu.cn}
\affiliation{Advanced Institute of Natural Sciences, Beijing Normal University, Zhuhai 519087, China}

\author{N. D. Ramesh Bhat}
\affiliation{International Centre for Radio Astronomy Research, Curtin University, Bentley, WA 6102, Australia}

\author{Yi Feng}
\affiliation{Research Center for Intelligent Computing Platforms, Zhejiang Laboratory, Hangzhou 311100, China}

\author{George Hobbs}
\affiliation{Australia Telescope National Facility, CSIRO Astronomy and Space Science, P.O. Box 76, Epping, NSW 1710, Australia}

\author{Richard N. Manchester}
\affiliation{Australia Telescope National Facility, CSIRO Astronomy and Space Science, P.O. Box 76, Epping, NSW 1710, Australia}

\author{Christopher J. Russell}
\affiliation{CSIRO Scientific Computing, Australian Technology Park, Locked Bag 9013, Alexandria, NSW 1435, Australia}

\author{R. M. Shannon}
\affiliation{Centre for Astrophysics and Supercomputing, Swinburne University of Technology, Hawthorn, VIC, 3122 Australia}
\affiliation{Australian Research Council Centre of Excellence in Graviational Wave Discovery (OzGrav)}

% \author{PPTA authors}
% \affiliation{PPTA affiiliations}
% \collaboration{The PPTA Collaboration}%\noaffiliation

%%%%%%%%%%%%%%%%%%%%%%%%%%%%%%%%%%%%%%%%%%%%%%%%%%%%%%%%%%%%%%%%%%%%%%%%%%%%%%%%
\begin{abstract}
Composed of ultralight bosons, fuzzy dark matter provides an intriguing solution to challenges that the standard cold dark matter model encounters on sub-galactic scales. The ultralight dark matter with mass $m\sim10^{-23} \rm{eV}$ will induce a periodic oscillation in gravitational potentials with a frequency in the nanohertz band, leading to observable effects in the arrival times of radio pulses from pulsars. Unlike scalar dark matter, pulsar timing signals induced by the vector dark matter are dependent on the oscillation direction of the vector fields. In this work, we search for ultralight vector dark matter in the mass range of $\unit[[2\times 10^{-24}, 2\times 10^{-22}]]{\rm{eV}}$ through its gravitational effect in the Parkes Pulsar Timing Array (PPTA) second data release. Since no statistically significant detection is made, we place $95\%$ upper limits on the local dark matter density as $\rho_{\rm{\tiny{VF}}} \lesssim \unit[5]{\rm{GeV/cm^{3}}}$ for $m\lesssim \unit[10^{-23}]{\rm{eV}}$. As no preferred direction is found for the vector dark matter, these constraints are comparable to those given by the scalar dark matter search with an earlier 12-year data set of PPTA.
\end{abstract}

	
\maketitle
	
%\keywords{gravitational waves -- instrumentation: detectors (PPTA) -- 
%		methods: data analysis} 
	
%%%%%%%%%%%%%%%%%%%%%%%%%%%%%%%%%%%%%%%%%%%%%%%%%%%%%%%%%%%%%%%%%%%%%%%%%%%%%%%%
\section{Introduction}

Numerous astrophysical observations, such as galaxy rotational curves \cite{1980ApJ...238..471R,1982ApJ...261..439R}, velocity dispersions \citep{1976ApJ...204..668F}, and gravitational lensing \citep{Massey:2010hh} reveal the existence of invisible matter, the so-called dark matter. In combination with observational evidence of the Universe’s accelerating expansion, the standard Lambda Cold Dark Matter ($\rm{\Lambda}$CDM) cosmological model has been established. Precision analyses of the cosmic microwave background show that dark matter constitutes $26\%$ of the total energy density of the present-day universe \citep{Planck:2018vyg}. 

The cold dark matter paradigm has achieved great success in describing the structure of galaxies on large scales \citep{Roszkowski:2017nbc, Marsh:2015xka,Wantz:2009it}, but it is met with puzzling discrepancies between the predictions and observations of galaxies and their clustering on small scales. For example, the N-body simulations based on the cold dark matter model show a much steeper central density profile in the dark matter halos than that inferred from the galaxy rotational curves (the ``core-cusp problem" \cite{Gentile:2004tb,deBlok:2009sp}). The predicted number of subhalos with decreasing mass grows much more steeply than what is observed around galaxies (the ``missing-satellites problem" \cite{Moore:1999nt,Klypin:1999uc}).

Because of the difficulty in solving the small-scale problems as well as the null result in searching for traditional cold dark matter candidates, e.g., weakly interactive massive particles \citep{Schumann:2019eaa}, alternative paradigms for dark matter have been proposed.
These include the warm dark matter \citep{Bode:2000gq} and fuzzy dark matter \citep{Hu:2000ke}.

The term ``fuzzy dark matter” often refers to ultralight scalar particles with a mass around  $m\sim \unit[10^{-22}]{\rm eV}$. Such a dark matter scenario can get the correct relic abundance through the misalignment mechanism similar to that of axions \citep{Fox:2004kb}; that is, when the initial value of the scalar field is away from its potential minimum, the field is condensed during inflation when its mass is smaller than the Hubble scale, and then starts a coherent oscillation as a non-relativistic matter at a later epoch. Fuzzy dark matter makes the same large-scale structure predictions as $\rm{\Lambda}$CDM, but the particle’s large de Broglie wavelength, $\lambda\sim \rm{kpc}$, suppresses the structure on small scales and thus explains well the corresponding smaller-scale observational phenomena \citep{Hui:2016ltb}.

Besides the scalar particle, a naturally light vector boson predicted in string-inspired models with compactified extra dimensions \citep{Goodsell:2009xc} can also act as a good fuzzy dark matter candidate. There are several mechanisms to produce vector dark matter with the correct relic abundance, such as the misalignment mechanism \citep{Nelson:2011sf,Nakayama:2019rhg}, quantum fluctuations during inflation \citep{Graham:2015rva, Nomura:2019cvc}, and decay of a network of global cosmic strings \citep{Long:2019lwl}. Because of their different spins, if the dark matter is assumed to have interaction with Standard Model, scalar and vector fields couple with Standard Model particles in ways which lead to different observable phenomena. For example, if the vector dark matter particle is a $U(1)_{B}$ (``$B$" refers to baryon ) or $U(1)_{B-L}$ (``$L$" refers to lepton) gauge boson, the so-called ``dark photon" would interact with ordinary matter \citep{Graham:2015ifn}, then it can be  detected with gravitational-wave interferometers because it exerts forces on test masses and results in displacements \citep{Pierce:2018xmy,LIGOScientificCollaborationVirgoCollaboration:2021eyz}; it can also be detected with binary pulsar systems via its effects on the secular dynamics of binary systems \citep{Blas:2016ddr,LopezNacir:2018epg}. Such a gauge effect is not applicable to scalar dark matter \citep{Antypas:2022asj}.

In addition to unknown interaction with the Standard Model, pure gravitational effects of fuzzy dark matter can also lead to observable results and help distinguish the scalar and vector dark matter. The dark matter field with ultralight mass has a wave nature with the oscillating frequency of $f_{\rm{dm}}=mc^2/h=\unit[2.4\times 10^{-9}(mc^2/10^{-23}\rm{eV})]{{Hz}}$. Such a coherently oscillating field leads to periodic oscillations in the gravitational potentials and further induces periodic signals with the frequency on the order of nanohertz \citep{Khmelnitsky:2013lxt,Nomura:2019cvc}, which falls into the sensitive range of the pulsar timing arrays (PTAs).
A PTA consists of stable millisecond pulsars for which times of arrival (ToAs) of radio pulses are monitored with high precision over a course of years to decades \citep{1978SvA....22...36S,Detweiler:1979wn,1990ApJ...361..300F}. Any unmodelled signal will induce timing residues, which represent the difference between the measured and predicted ToAs.

In contrast to the ultralight scalar dark matter, the timing residuals caused by the ultralight vector dark matter are dependent on the oscillation direction of the vector fields \citep{Khmelnitsky:2013lxt, Nomura:2019cvc}.
Several previous works have used PTA data to search for ultralight scalar dark matter \citep{Porayko:2014rfa, Porayko:2018sfa, Kato:2019bqz}.
In a recent work \cite{PPTA:2021uzb},
a search was performed for the dark photon dark matter in the PPTA second data release (DR2) based on the gauge effect.
This resulted in upper limits on the coupling strength between dark photons and ordinary matter, assuming that all dark matter is composed of ultralight dark photons.
In this work, we search for ultralight vector dark matter in the mass range of $\unit[[2\times 10^{-24}, 2\times 10^{-22}]]{\rm{eV}}$ in the PPTA DR2 data set based on the gravitational effect without assuming its interaction with Standard Model particles.

This paper is organized as follows. In Sec. \ref{sec2} we describe the observable pulsar timing effects induced by vector dark matter.
In Sec. \ref{sec3} we provide details of our data analysis.
We present the results and conclusions in Sec. \ref{sec4}. In the following sections, we set $c=\hbar=1$.

%%%%%%%%%%%%%%%%%%%%%%%%%%%%%%%%%%%%%%%%%%%%%%%%%%%%%%%%%%%%%%%%%%%%%%%%%%%%%%%%  
\section{Gravitational Effect from vector dark matter}\label{sec2}

In this section, we first introduce the timing residuals caused by the gravitational effect from the ultralight vector dark matter in the Galaxy; a more detailed derivation can be found in Ref. \citep{Nomura:2019cvc}.
 
Assuming no coupling between the ultralight particles and any other fields, we take the action for a free vector field $A_{\mu}$ with mass $m$ as,
\e
S = \int d^4x \sqrt{-g} \(-\frac{1}{4} F^{\mu\nu}F_{\mu\nu} - \frac{1}{2}m^2 A_{\mu}A^{\mu}\),
\q
where $g$ is the determinant of the metric $g_{\mu\nu}$ and $F_{\mu\nu}=\partial_{\mu}A_{\nu}-\partial_{\nu}A_{\mu}$.
On Galactic scales, the cosmic expansion is negligible and the background is approximately Minkowski. The energy-momentum tensor carried by the vector dark matter induces perturbations into the metric which, in the Newtonian gauge, can be written as
\e
\begin{split}
ds^2 = &\eta_{\mu\nu}dx^{\mu}dx^{\nu}-2\Phi(t,\bx)dt^2 + 2\Psi(t,\bx)\dt_{ij}dx^{i}dx^{j} \\
       &+h_{ij}(t,\bx)dx^{i}dx^{j}\, ,
\end{split}
\label{metric}
\q
where $\eta_{\mu\nu}=\rm{diag}(-1,+1,+1,+1)$ is the background Minkowski metric, $\Phi$ and $\Psi$ are gravitational potentials, and $h_{ij}$ describes the traceless spatial metric perturbations. $h_{ij}$ is absent in the scalar-field case and demonstrates the anisotropy induced by additional degrees of freedom in vector fields.

With a huge occupation number, the vector field can be described as a classical wave with a monochromatic frequency determined by its mass.
This is a good approximation because the characteristic speed of the dark matter is non-relativistic $v\sim10^{-3}$. During inflation, only the longitudinal mode of the vector fields survives \citep{Graham:2015rva}, so the equation of motion of the vector field is given by the component in the oscillating direction $\hat{k}=(\sin\theta \cos\phi, \sin\theta \sin\phi, \cos\theta)$,
\e\label{A}
A_{\hat{k}}(t,\bx) = A(\bx) \cos(mt+\al(\bx)).
\q
The vector fields contribute a time-independent energy density
\e\label{rho}
\rho_{\rm{\tiny{VF}}}(x)=\frac{1}{2}m^2 A^2(x),
\q
and an anisotropic oscillating pressure which leads to oscillating gravitational potentials. By solving the photon geodesic equation from the pulsar to the Earth under the metric \Eq{metric}, it is found that the metric perturbations that give rise to the observable effects in PTAs are from the spatial components (see the Appendix of \citep{Nomura:2019cvc}). Furthermore, splitting the potential $\Psi$ into a dominant time-independent part and an oscillating part and solving the linear Einstein equation by neglecting the spatial gradient of the oscillating part (which is suppressed by order of $v\sim 10^{-3}$), the spatial perturbations take the following form, 
% In particular, the oscillating parts in the metric that give rise to the observable effects in PTAs are from the spatial components, i.e.,
% \begin{small}
\m
\!\!\!\!\Psi(t,\bx)&\!=\!&\Psi_0(\bx)+\Posc(\bx)\cos(2mt+2\al(\bx)),\\
\!\!\!\!h_{ij}(t,\bx)&\!=\!&\hosc(\bx)\!\cos(2mt\!+\!2\al(\bx))(\hat{l}\otimes \hat{ l}\!+\! \hat{n}\otimes \hat{n}\!-\! 2\hat{k}\otimes \hat{k} ),
\n
% \end{small}
where $\Psi_0(x)$ is the potential independent of time determined by the local energy density of the vector field dark matter $\rho_{\rm{\tiny{VF}}}(x)$, and $\hat{l}$ and $\hat{n}$ are the unit vectors perpendicular to the propagation direction given by $\hat{l}=(\sin\phi, -\cos\phi, 0)$ and $\hat{n}=(\cos \theta \cos \phi, \cos\theta\sin\phi, -\sin\theta)$, respectively. 
% \textbf{Solving the linear Einstein equation by overlooking the spatial gradients that is suppressed by the order of  $v\sim 10^{-3}$ compared with the time derivative }
The amplitudes of potentials in the oscillation part can also be related to $\rho_{\rm{\tiny{VF}}}(x)$ through the relationship between $A(x)$ with $\rho_{\rm{\tiny{VF}}}(x)$ (\Eq{rho}),
\m
\Psi_{\mathrm{osc}}(\mathbf{x}) &=& -\frac{\pi G \rho_{\rm{\tiny{VF}}}(\mathbf{x})}{3 m^{2}} \notag\\
&=&-2.2 \times 10^{-16} \frac{\rho_{\rm{\tiny{VF}}}(\mathbf{x})}{\rho_0} \left(\frac{10^{-23} \mathrm{eV}}{m}\right)^{2}, \\
h_{\mathrm{osc}}(\mathbf{x}) &=&\frac{8 \pi G \rho_{\rm{\tiny{VF}}}(\mathbf{x})}{3 m^{2}} \notag\\
&=&1.7 \times 10^{-15} \frac{\rho_{\rm{\tiny{VF}}}(\mathbf{x})}{\rho_0} \left(\frac{10^{-23} \mathrm{eV}}{m}\right)^{2}. 
\label{h_rho}
\n
where we use the measured local energy density $\rho_0=\unit[0.4]{ \mathrm{GeV} / \mathrm{cm}^{3}}$ \citep{Salucci:2010qr} as the normalized factor, 
and the oscillation frequency in potentials is given by:
\e
f=\frac{2 m}{2 \pi}=4.8 \times 10^{-9} \mathrm{~Hz}\left(\frac{m}{10^{-23} \mathrm{eV}}\right).
\q
The oscillating part of spatial metric  $\Psi$ and $h_{ij}$ induces a mearsurable redshift in the radio pulse propagating from a pulsar to the Earth
\m
z_{\Psi}(t)=&&\Psi_{\mathrm{osc}}(\bx_{e})\cos(2mt+2\al(\bx_{e}))\notag \\
&&-\Psi_{\rm{osc}}(\bx_{p})\cos[2m(t-|\bx_{p}|)+2\al(\bx_{p})],\\
z_h(t)=&&\frac{1}{2}\hat{p}^i \hat{p}^j [h_{ij}(t,\bx_{e})-h_{ij}(t-|\bx_{p}|,\bx_{p})],
\n
where $\bx_{e}$ and $\bx_{p}$ respectively represent the location of the Earth and the pulsar, and $\hat{p}$ is the unit vector pointing to the pulsar. As the distance between most pulsars and the Earth is at the order of $\mathcal{O}(\mathrm{kpc})$, which is comparable to the de Broglie wavelength of the dark matter, it is legitimate to assume that they are in a region where the vector dark matter keeps its coherent oscillation direction, and that the Earth term and the pulsar term take the same amplitudes $\Psi_{\rm osc}$ and $h_{\rm osc}$.

Integrating the redshifts separately, the results combine into the total timing residuals,
 \m
 R_t=&&R_{\Psi}+R_{h}\notag\\
  =&&\frac{h_{\rm osc}}{\pi f}\left\{\frac{1}{2}\left[(\hat{p}\cdot \hat{l})^2+(\hat{p} \cdot \hat{n})^2-2(\hat{p} \cdot \hat{k})^2\right]-\frac{1}{8}\right\}\notag \\
  && \times \sin (\alpha_{e}-\alpha_{p}) \cos \left(2 \pi f t+\alpha_{e}+\alpha_{p}\right),
  \label{R_t}
 \n
 where we have defined the phases in Earth term and pulsar term as $\alpha_{e}=\alpha(\bx_{e})$ and $\alpha_{p}=\alpha\left(\bx_{p}\right)-m\left|\mathbf{x}_{p}\right|$, respectively.
\Eq{R_t} shows that timing residuals induced by the coherent oscillation of vector dark matter is angle dependent, which is a distinctive feature in comparison to the scalar dark matter; see Ref. \citep{Nomura:2019cvc} for a more detailed comparison. 

%%%%%%%%%%%%%%%%%%%%%%%%%%%%%%%%%%%%%%%%%%%%%%%%%%%%%%%%%%%%%%%%%%%%%%%%%%%%%%%%
\section{Data analysis}\label{sec3}

Now we turn to search for ultralight vector dark matter in the PPTA DR2 data set, which includes observations for 26 pulsars with a timespan up to 15 years. 
% \textbf{However, the Bayesian data analysis that we will adopt to conduct the search is difficult to perform in general due to the quite large parameter space (over 800 parameters for the current PPTA data set for the signal and the noise model following \citep{Goncharov:2020krd}, see details below). And unlike a broadband gravitational-wave background whose sensitivity is largely affected by the number of pulsars, the sensitivity of a narrowband signal is mostly determined by the timing residuals of the individual pulsars \citep{Siemens_2013}.}
By balancing sensitivity and computational costs, we choose the six best pulsars, i.e., those with relatively long observational timespan and high timing precision in the array. A summary of the basic properties of these six pulsars is given in \Table{properties}.

\begin{table}[!htbp]
	\caption{Basic properties of the 6 pulsars used in our analysis.  RMS -- the weighted root-mean-square band-averaged post-fit timing residuals,  $N_{\rm{obs}}$ -- the number of observations, $N_{\rm{ToA}}$ -- the number of ToAs, Span -- observational data span. See Ref. \citep{Kerr:2020qdo} for details.}
	\label{properties}
	%    \centering
	\begin{tabular}{c c c c c}
		\hline
		\hline
		Pulsar Name &RMS\,[$\mu$s] &$N_{\rm{obs}}$&$N_{\rm{ToA}}$ &Span\,[$\yr$] \,\\
		\hline
		J0437$-$4715 &0.59&4149&29262&15.0\\
		\hline
		J1600$-$3053&0.58&1096&7047&14.2\\
		\hline
		J1713+0747&0.32&1049&7804&14.2\\
		\hline
		J1744$-$1134&0.46&939&6717&14.2\\
		\hline
		J1909$-$3744&0.24&2223&14627&14.2\\
		\hline
		J2241$-$5236&0.26&821&5224&8.2\\
		\hline
	\end{tabular}
	\vspace{2ex}
\end{table} 

We process the data in the same way as Refs. \cite{Porayko:2018sfa,PPTA:2021uzb}. To extract the target signal from the ToAs, one needs to provide a comprehensive analysis on the noise that might be present in timing residuals. After subtracting the expected arrival times described by the timing model, the timing residuals can be decomposed into 
\e
\delta \boldsymbol{t}= M {\boldsymbol{\epsilon}} + \delta \boldsymbol{t}_{n}+  \delta \boldsymbol{t}_{s},
\q
where $M {\boldsymbol{\epsilon}}$ accounts for the inaccuracy of the timing model with $M$ being the design matrix and $\boldsymbol{\epsilon}$ being the vector of timing model parameter offsets, $\delta \boldsymbol{t}_{n}$ contains noise contributions, and $ \delta \boldsymbol{t}_{s}$, given by \Eq{R_t}, is the signal that we are searching for.

\begin{table*}[!htbp]
	%    \centering
	\footnotesize
	\caption{Parameters and their prior distributions used in the analyses. Here U and log-U represent a uniform and log uniform distribution, respectively. Here ``one parameter for PTA" means the parameter is common in the whole data set, while ``one parameter per pulsar" indicates the parameter varies from pulsar to pulsar; the same goes for the case of  ``one parameter per band/system" and ``one parameter per exponential-dip event".}
	\label{prior}
	\begin{tabular}{c c c c}
		\hline
		\textbf{Parameter} & \textbf{description} & \textbf{prior} & \textbf{comments} \\
		\hline
		\multicolumn{4}{c}{White noise}\,\\	        
		$E_{k}$ & EFAC per backend/receiver system & $\uni[0, 10]$ & single-pulsar analysis only \\
		$Q_{k}\,[\mrm{s}]$ & EQUAD per backend/receiver system & $\logu[-8.5, -5]$ & single-pulsar analysis only \\
		$J_{k}\,[\mrm{s}]$ & ECORR per backend/receiver system & $\logu[-8.5, -5]$ & single-pulsar analysis only \\
		\hline
		\multicolumn{4}{c}{Red noise (including SN and DM)} \\
		$\Am_{\RN}$ & Red-noise power-law amplitude &$\logu[-20, -8]$ & one parameter per pulsar\, \\
		$\gamma_{\RN}$ &red-noise power-law index  &$\uni[0,10]$ & one parameter per pulsar\, \\
		\hline
		\multicolumn{4}{c}{Band/System noise}\,\\
		$\Am_{\BN,\GN}$ & band/group-noise power-law amplitude &$\logu[-20, -8]$ & one parameter per band/system\, \\
		$\gamma_{\BN,\GN}$ &band/group-noise power-law index &$\uni[0,10]$ &one parameter per band/system\, \\
		\hline
		\multicolumn{4}{c}{Deterministic event}\,\\
		$\Am_{\mathrm{E}}$ & exponential-dip amplitude &$\logu[-10, -2]$ & one parameter per exponential-dip event \, \\
		$ t_{\mathrm{E}}\, [\mrm{MJD}]$ &time of the event &
		$\uni[54500, 54900]$ for PSR J1713 & first exponential-dip event\\
		$ $ & &$\uni[57500, 57520]$ for PSR J1713 & second exponential-dip event\\
		$\mrm{log_{10}} \tau_{\mrm{E}}\, [\mrm{MJD}]$ & relaxation time for the dip &$\uni[\mrm{log_{10}}5, 2]$ &one parameter per exponential-dip event \, \\
		\hline
		\multicolumn{4}{c}{Common noise} \\
		$\Am_{\rm{CN}}$ & common-noise power-law amplitude &$\logu[-18, -11]$ & one parameter for PTA\, \\
		$\gamma_{\rm{CN}}$ &common noise power-law index  &$\uni[0,7]$ & one parameter for PTA\, \\
		\hline
		\multicolumn{4}{c}{Ultralight vector dark matter signal}\,\\
		$h_{\rm{osc}}$ & oscillation amplitude &$\logu[-19, -10]$ (search) & one parameter for PTA\, \\
		&&$\uni[-19,-10]$ (limit)\,\\
		$\al_e$ &oscillation phase on Earth &$\uni[0,2\pi]$ & one parameter for PTA\, \\
		$\al_p$ & equivalent oscillation phase on pulsar &$\uni[0,2\pi]$ & one parameter per pulsar\, \\
		$\cos\theta$ & polar angle of propagation direction&$\uni[-1, 1]$ & one parameter for PTA\, \\
		$\phi$ &  azimuth angle of propagation direction &$\uni[0,2\pi]$ & one parameter for PTA\, \\
		$f\,[\mrm{Hz}]$ & oscillation frequency &$\logu[-9,-7]$ (search) & one parameter for PTA\, \\
		& &delta function (limit) & fixed\,\\
		\hline
		\multicolumn{4}{c}{BayesEphem}\,\\
		$z_{\mrm{drift}}$ & drift-rate of Earth’s orbit about ecliptic z-axis &$\uni[-10^{-9}, 10^{-9}]$ & one parameter for PTA\, \\
		$\Dt M_{\mrm{jupiter}}\,[\Msun]$ &perturbation to Jupiter’s mass  & $\mcN(0, 1.55\times10^{-11})$ &one parameter for PTA \, \\
		$\Dt M_{\mrm{saturn}}\,[\Msun]$ &perturbation to Saturn’s mass  & $\mcN(0, 8.17\times10^{-12})$ &one parameter for PTA \, \\
		$\Dt M_{\mrm{uranus}}\,[\Msun]$ &perturbation to Uranus’s mass  & $\mcN(0, 5.72\times10^{-11})$ &one parameter for PTA \, \\
		$\Dt M_{\mrm{neptune}}\,[\Msun]$ &perturbation to Neptune’s mass & $\mcN(0, 7.96\times10^{-11})$ &one parameter for PTA \, \\
		$\mrm{PCA}_i$ &principal components of Jupiter’s orbit&$\uni[-0.05,0.05]$ &six parameters for PTA \, \\
		\hline
	\end{tabular}
% 	\vspace{2ex}
\end{table*}

The noise in all of these pulsars from possible stochastic and deterministic processes has been analyzed by Ref.~\cite{Goncharov:2020krd} in great detail. The stochastic noise processes contain white noise and time-correlated red noise. White noise accounts for measurement uncertainties; they are modeled by three parameters EFAC, EQUAD, ECORR \citep{NANOGrav:2015qfw}, with EFAC being the scale factor of ToA uncertainty,  EQUAD being an extra component independent of uncertainty and ECORR being the excess variance for sub-banded observations. The red noise includes the spin noise (SN; \citep{Shannon:2010bv}) from rotational irregularities of the pulsar itself, the dispersion measure variations \citep{Keith:2012ht} due to the change in column density of ionized plasma in the interstellar medium, and the band noise (BN) and system (``group") noise (GN) that are only present in a specific band or system \citep{Lentati:2016ygu}. Red noise is modeled by a power-law spectrum with the amplitude parameter $\Am_{\RN}$ and spectral index $\gamma_{\RN}$.  In our analysis, we set the number of Fourier frequencies $N_{\rm{mode}}=30$ following Ref.~\citep{Arzoumanian:2020vkk} in the calculation of the covariance matrix. For deterministic noise contributions, a typical example is the exponential dip that might be attributed to the sudden change in dispersion in the interstellar medium \citep{Lentati:2016ygu,Keith:2012ht} or change in pulse profile shape \citep{shannon1643}, and it can be described by an exponential function.

Meanwhile, some systematic errors should be taken into consideration. We use the \texttt{BayesEphem} module \citep{NANOGrav:2020tig} to account for potential uncertainties in the solar system ephemeris (SSE); we adopt JPL DE438 \citep{DE438} to project ToAs from the local observatory to the solar system barycenter. Moreover, the North American Nanohertz Observatory for Gravitational Waves (NANOGrav), PPTA, the European Pulsar Timing Array (EPTA), and the International Pulsar Timing Array (IPTA) collaborations all report evidence for an uncorrelated common process (UCP) which can be modeled by a power-law spectrum in their lastest data sets \citep{Arzoumanian:2020vkk, Goncharov:2021oub,Antoniadis:2022pcn,Chen:2021rqp}. Although no definite evidence was found for a Hellings-Downs correlation which is deemed to be necessary for the detection of stochastic gravitational-wave background, the presence of UCP is taken as a promising sign of the gravitational-wave background \citep{Arzoumanian:2020vkk}. Positive Bayesian evidence supporting a scalar-transverse correlation in the process was found in some publications \citep{Chen:2021ncc,Chen:2021wdo}. However, simulations based on the PPTA DR2 showed that even when no signal is present, the UCP pops out when pulsars have similar intrinsic timing noise \citep{Goncharov:2021oub}. Despite the continuing efforts \citep{Xue:2021gyq,Wu:2021kmd,Chen:2022azo,Bian:2022tju}, the nature of the UCP remains to be determined, and we hence treat it as a common noise in the analysis.

As \Eq{R_t} indicates, the vector dark matter signal that we are searching for is described by six parameters: the oscillation amplitude $h_{osc}$, the oscillation frequency $f$, the oscillation (propagation) direction described by the polar and azimuth angles ($\theta, \phi$) and the equivalent phase term in pulsar $\al_{p}$ and in the Earth $\al_{e}$. In the analyses, we first perform the parameter estimations by including the white noise, red noise, band/system noise, and deterministic noise following \citep{Goncharov:2020krd} for each single pulsar. Then we collect all the chosen pulsars as a whole PTA and allow noise parameters to vary simultaneously with the signal parameters. As the white noise parameters should have little or no correlation with the dark matter parameters, we fix the white noise parameters to their maximum-likelihood values from the single pulsar noise analyses. Fixing white noise parameters should have negligible impact on our results \citep{NANOGRAV:2018hou} but significantly reduce the computational cost. All the parameters and their priors are listed in \Table{prior}.

Similar to the procedure of searching for a gravitational wave background \citep{Arzoumanian:2020vkk, Goncharov:2021oub}, we perform Bayesian inference to extract information from the data. 
First, we need to determine whether the data $\Dm$ supports the existence of the signal by calculating the Bayes factor between the noise-plus-signal hypothesis $\Hm_1$ and the noise-only hypothesis $\Hm_0$,
\e
\rm{BF}=\frac{\rm{Pr}(\Dm|\Hm_1)}{\rm{Pr}(\Dm|\Hm_0)}\, .
\q
Here $\rm{Pr}(\Dm|\Hm)$ denotes the evidence given by the integral of the product of the likelihood $\mathcal{L}$ and the prior probability  $\pi$ over the prior volume,
\e
\rm{Pr}(\Dm|\Hm)=\int \mathcal{L}(\Dm|{\Theta})\pi({\Theta}) d^n{\Theta},
\q
where $n$ is the dimension of the parameters $\Theta$.
If we do not find significant evidence for the target signal, we place constrains on certain parameters. In the work, we derive the $95\%$ upper limit for the oscillation amplitude $\bar{h}_{\rm osc}$ from its marginal posterior probability distribution,
% \m
% 0.95=\int_0^{\bar{h}_{\rm{osc}}}\pi(h_{\rm{osc}})d\, h_{\rm{osc}}\int \mathcal{L}(\Dm,\Theta)\pi(\Theta')d^{n-1}\Theta' ,
% \n
\e
0.95=\frac{1}{\rm{Pr}(\Dm|\Hm)}\int_0^{\bar{h}_{\rm{osc}}} dh_{\rm{osc}}\int \mathcal{L}(\Dm|\Theta)\pi(\Theta)d^{n-1}\Theta' ,
\q
where $\Theta'$ denotes all the other parameters except $h_{\rm{osc}}$.
We use the \texttt{enterprise} \citep{enterprise} and \texttt{enterprise\_extension} \citep{enterprise_extensioins} packages to evaluate the likelihood and compute the Bayes factor using the product-space method \citep{10.2307/2346151,Hee:2015eba,Taylor:2020zpk}. For the Markov-chain Monte-Carlo sampling needed for parameter estimation, we employ the \texttt{PTMCMCSampler} package \citep{justin_ellis_2017_1037579}.

%%%%%%%%%%%%%%%%%%%%%%%%%%%%%%%%%%%%%%%%%%%%%%%%%%%%%%%%%%%%%%%%%%%%%%%%%%%%%%%%

%  However, from the posterior distribution, we can see a peak near $\log_{10}f\sim -7$ in the parameter space of the frequency. Then we narrow the frequency range in $[10^{-7.2},10^{-7.0}]$, and get the peak frequency more accurately at $\log_{10}f\sim -7.04$,  we further fix the frequency parameter to the peak value and find the Bayes factor has a significant enhancement to be near 30.
% Similar result has been reported in both \cite{Porayko:2018sfa} and \cite{PPTA:2021uzb}, and the source for the signal is not clear at present.
\section{results and discussion}\label{sec4}

We first determine whether there is a vector dark matter signal in the data by comparing the signal hypothesis $\mathcal{H}_1$ and the null hypothesis $\mathcal{H}_0$. The log Bayes factor, $\ln \rm{BF}$, is about 13.0 for the parameter range $\log_{10}(f/\mathrm{Hz}) \in [-7.2,-7.0]$, seemingly implying strong evidence for a signal. However, when we conduct the search in two separate logarithmic frequency bands $[-9.0,-7.0]$ and $[-7.2,-7.0]$ by excluding PSR J0437$-$4715, the $\ln \rm{BF}$ is found to be 0.5 and 1.5, respectively; these are small Bayes factors that indicates no preference for the signal hypothesis. Therefore, the suspected ``signal" is completely due to PSR J0437$-$4715, and we conclude that it is not a genuine signal. A similar result has been reported in Ref.~\cite{PPTA:2021uzb}.

\begin{figure}[htbp!]
	\centering
	\includegraphics[width=0.5\textwidth]{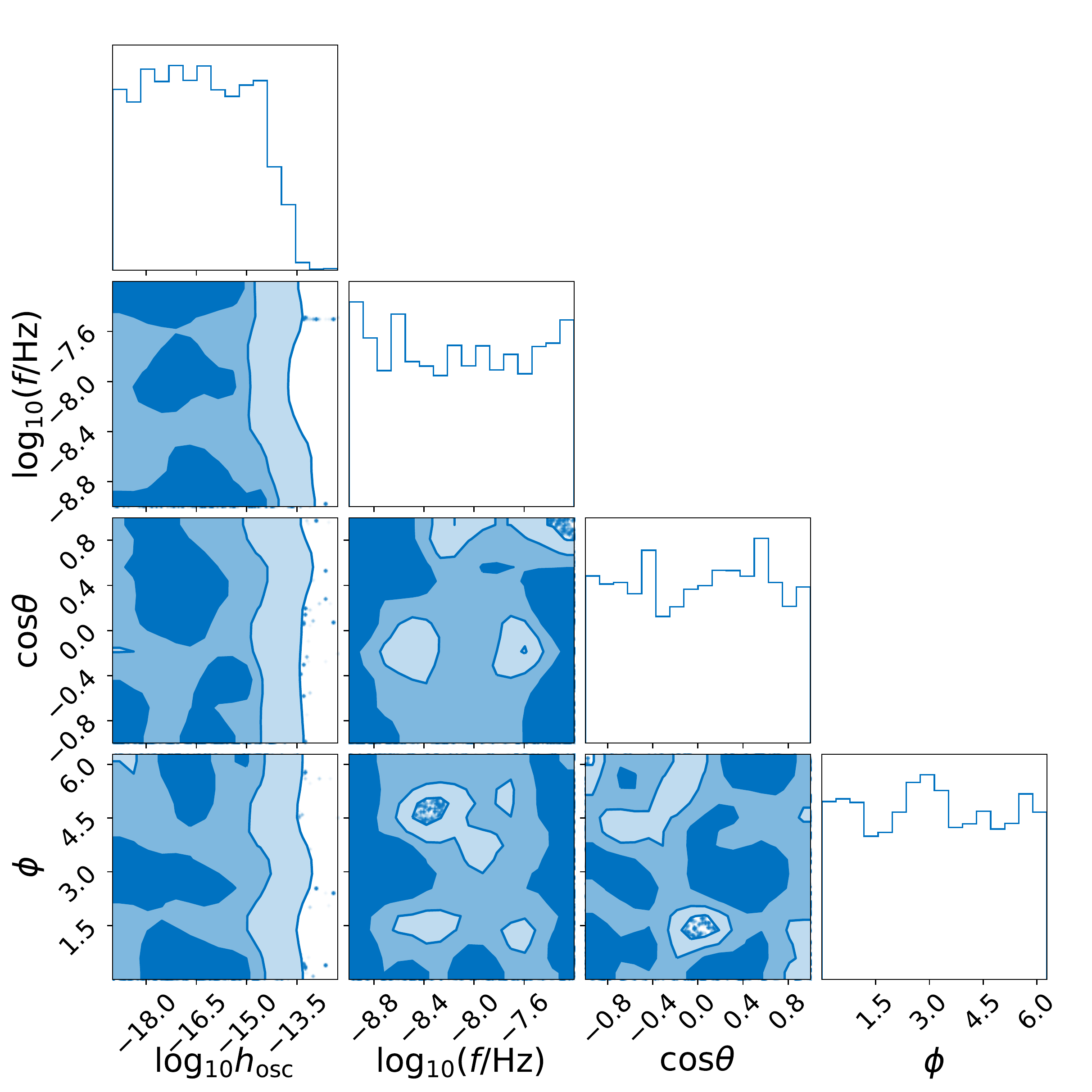}\caption{ \label{freefreq}The posterior distributions for the parameters of the vector dark matter signal, containing the oscillation amplitude $h_{\rm{osc}}$, oscillation frequency $\log_{10}(f/\mathrm{Hz})\in [-9.0,-7.2]$, and the angles of propagation direction $\theta$ and $\phi$. These posteriors have been marginalized over oscillation phase parameters $\alpha_{e}$ and $\alpha_p$.}
\end{figure}

\begin{figure}[htbp!]
	\centering
	\includegraphics[width=0.48\textwidth]{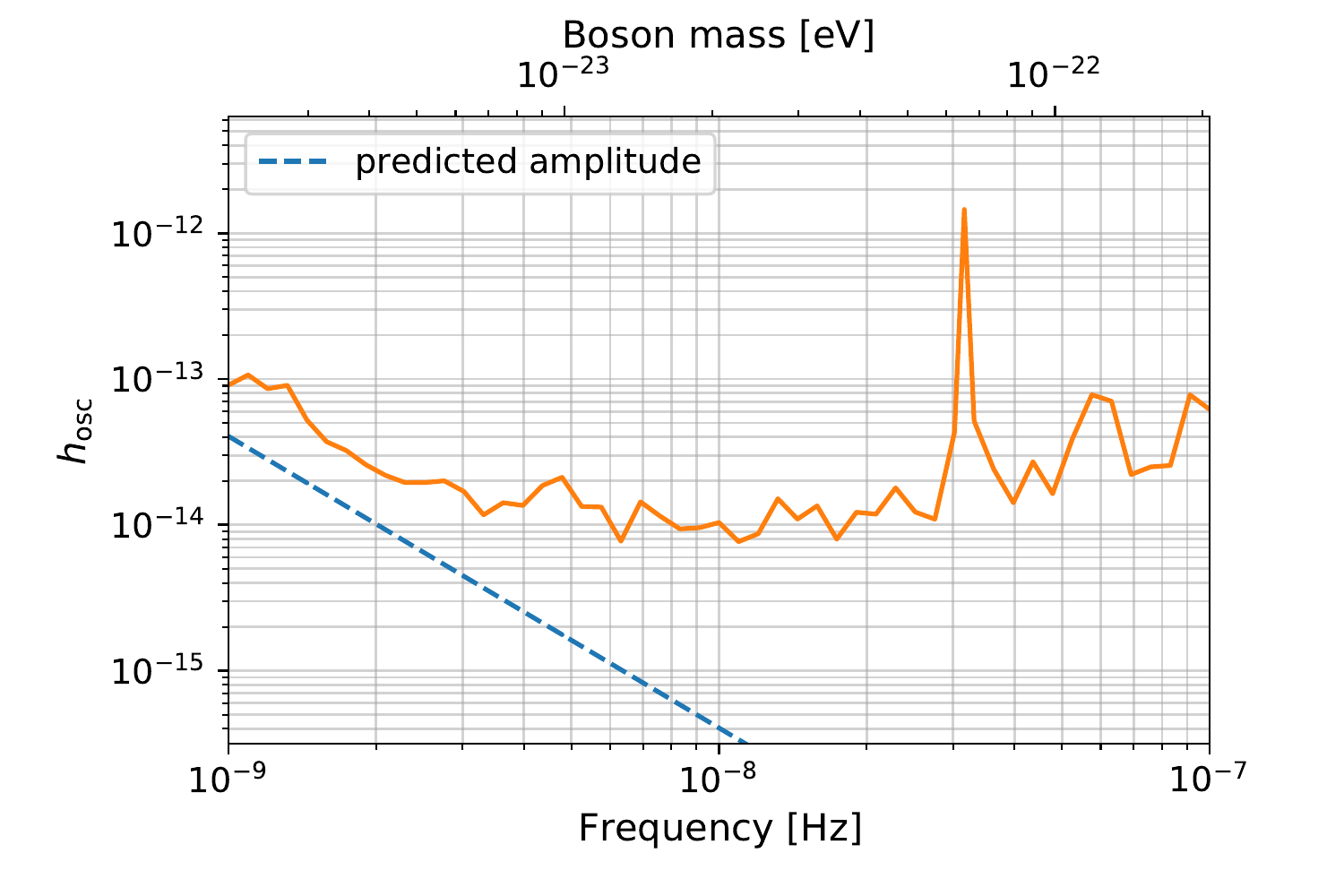}\caption{\label{hosc_f}Upper limits on the signal amplitude $h_{\rm{osc}}$ (orange line), generated by the vector dark matter in the Galatic as a function of frequency (mass). The blue dashed line shows the model amplitude $h_{\rm{osc}}$, assuming $\rho_{\rm{\tiny{VF}}}=\unit[0.4]{ \rm{GeV/cm^{3}}}$, given by \Eq{h_rho}. The peak in the orange curve is due to the loss in sensitivity at the frequency $1\rm{yr}^{-1}$ caused by fitting for the pulsar's sky location and proper motion \citep{Zhu:2014rta}.}
\end{figure}

\begin{figure}[htbp!]
	\centering
	\includegraphics[width=0.5\textwidth]{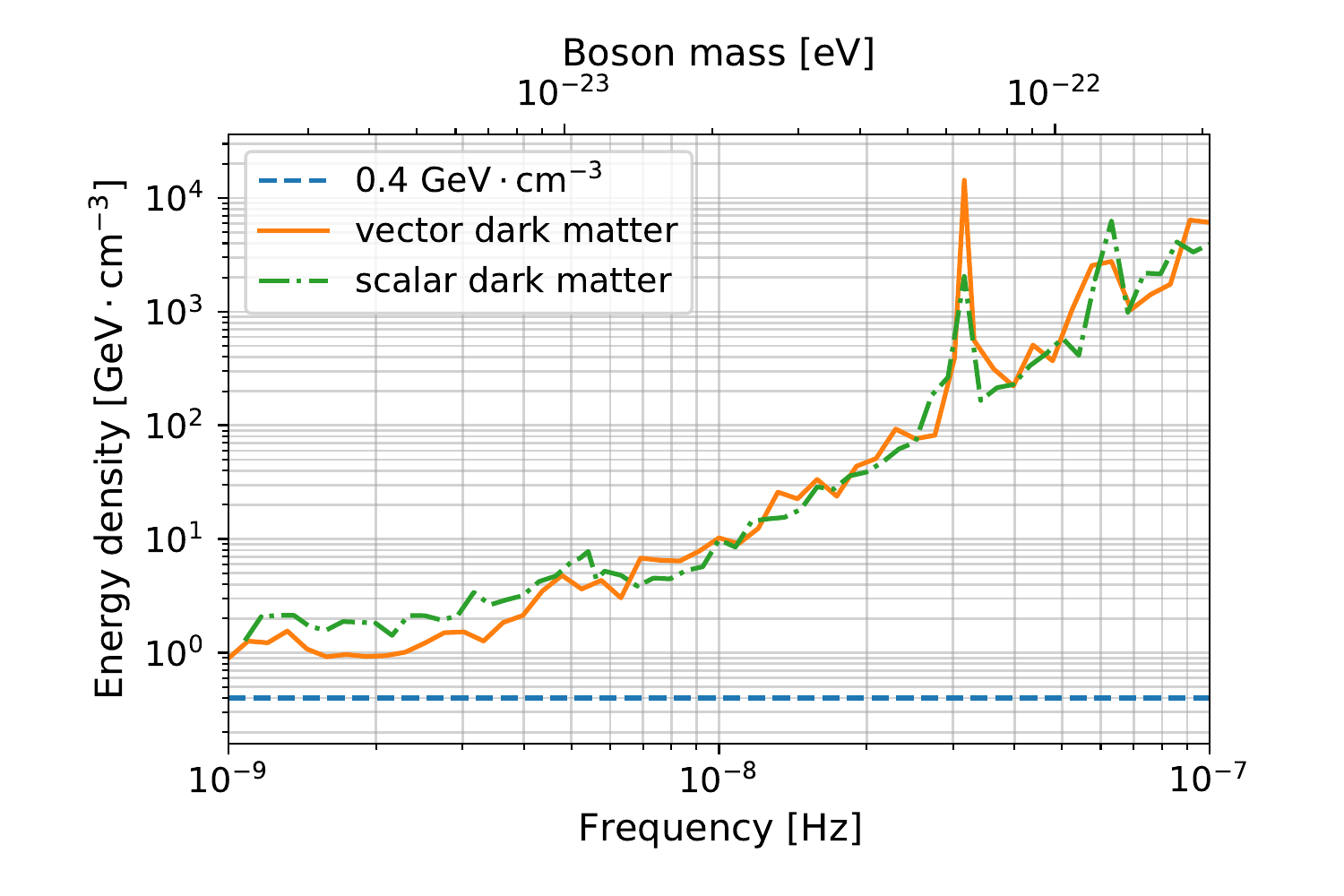}\caption{\label{rho_f} Upper limits on the vector dark matter density $\rho$ in the vicinity of the Earth (orange line). The green dot-dashed line denotes the upper limits on the scalar dark matter density from PPTA 12-year data set \cite{Porayko:2018sfa} and the blue dashed line denotes the measured dark matter density of $\rho_0=\unit[0.4]{ \rm{GeV/cm^{3}}}$.}
\end{figure}

\Fig{freefreq} shows the posterior distributions of the vector dark matter signal parameters after marginalizing over oscillation phases $\alpha_{e}$ and $\alpha_p$.
Since we find no evidence of a vector dark matter signal, these posterior distributions resemble those of priors except that we are able to exclude large oscillation amplitudes, i.e., $h_{\rm{osc}} \lesssim 10^{-13.5}$.
In \Fig{hosc_f}, we show the $95\%$ upper limits on the oscillation amplitude 
$h_{\rm osc}$ as a function of frequency. We also plot the predicted amplitude according to \Eq{h_rho} by taking the vector field dark matter density $\rho_{\rm{\tiny{VF}}}$ equal to the measured local dark matter density $\rho_0=\unit[0.4]{\rm{GeV/cm^{3}}}$, assuming that all the dark matter is composed of ultralight vector fields. The predicted amplitude is always below the upper limits, implying we cannot exclude the possibility that the vector fields with masses considered in this work constitute all the dark matter. A more intuitive picture can be obtained by translating the amplitude into the vector field dark matter energy density $\rho_{\rm{\tiny{VF}}}$ through \Eq{h_rho} and placing the $95\%$ upper limit on the energy density, as shown in \Fig{rho_f}.
% As the amplitude can be related to the dark matter energy density $\rho_{\rm{\tiny{VF}}}$ through \Eq{h_rho} , we also place the $95\%$ upper limit on the energy density, as shown in \Fig{rho_f}.
As a comparison, we also plot the upper limits derived from a scalar dark matter search presented in Ref.~\cite{Porayko:2018sfa}.
We note that the lighter mass provides tighter constraints in our search range. The strongest bound on the dark matter energy density is $\unit[1]{ \rm{GeV/cm^3}}$ at the lowest frequency $10^{-9}$ Hz. This is above the local energy density of $\rho_0=\unit[0.4]{ \rm{GeV/cm^3}}$, so we cannot place effective constraints on the mass of vector dark matter from current PTA data sets.

The results from this work are consistent with the scalar dark matter search performed with an earlier version of the PPTA data published in 2018.
Specifically, our constraints on vector dark matter density $\rho_{\rm{\tiny{VF}}}<\unit[5]{\rm{GeV/cm^{3}}}$ for $m\lesssim \unit[10^{-23}]{\rm{eV}}$ is only slightly more stringent than the scalar dark matter density limit of $\unit[6]{\rm{GeV cm^{-3}}}$ given by \citep{Porayko:2018sfa}.
This is unsurprising because only when the vector field oscillates along a particular direction that is parallel to the line of sight to the pulsar, can we expect the timing residuals induced by the gravitational effect to be three times that caused by the scalar dark matter. However, we do not find a preferred oscillation direction of vector dark matter (see \Fig{freefreq}) in the data set.

Several cosmological and astrophysical probes can place constraints on the mass of ultralight dark matter when considering the model has only gravitational coupling. Planck cosmic microwave background data implies a bound on the ultralight dark matter energy density fraction for the $10^{-33} - \unit[10^{-24}]{\rm{eV}}$ mass range \cite{Hlozek:2017zzf}. The Lyman-$\alpha$ forest used to trace the underlying dark matter distribution at high redshifts excludes the possibility that the ultralight particles with mass lighter than $\unit[2 \times 10^{-20}]{ \rm{eV}}$ make up all the dark matter \citep{Rogers:2020ltq}. While the above constraints come from axions and their applicability to ultralight vector dark matter needs to be discussed, the black hole superradiance estimates from supermassive black hole spin measurements constrain the vector dark matter within the mass range $6\times 10^{-20}$--$\unit[2 \times 10^{-17}]{ \rm{eV}}$  \citep{Baryakhtar:2017ngi}.  Although there is no consensus on the constraints on the mass of ultralight dark matter because most of the experiments are subject to their own uncertainties, it is important to note that PTA experiments can constrain the energy density of ultralight dark matter independently and determine whether dark matter is dominated by the ultralight vector particles and thus provide complementary tools to other experiments.
 
Looking into the future, PTAs based on the Five-hundred-meter Aperture Spherical Telescope \citep{Nan:2011um}, MeerKAT \cite{meerkat},  and Square Kilometer Array (SKA) \citep{Lazio_2013} with wider frequency bands and large collecting areas, will increase the sensitivity significantly. In the SKA era, a conservative estimate is that 10 years of observations for the 10 best pulsars with an observing cadence of once every 14 days have the potential to constrain the contribution of ultralight dark matter down to $10\%$ of the local dark matter density for $m<\unit[10^{-23}]{\rm{eV}}$ \citep{Porayko:2018sfa}.
In the shorter terms sensitivity improvments could be achieved through searches with the IPTA.
Combining efforts from PTAs and other experiments will greatly advance our understanding of the nature of dark matter by studying a wide range of dark matter models.

%\newpage   
%%%%%%%%%%%%%%%%%%%%%%%%%%%%%%%%%%%%%%%%%%%%%%%%%%%%%%%%%%%%%%%%%%%%%%%%%%%%%%%%
%%%%%%%%%%%%%%%%%%%%%%%%%%%%%%%% acknowledgements %%%%%%%%%%%%%%%%%%%%%%%%%%%%%%
\begin{acknowledgments}
We thank the referee for very useful comments.
We acknowledge the use of HPC Cluster of ITP-CAS and HPC Cluster of Tianhe II in National Supercomputing Center in Guangzhou. QGH is supported by the National Key Research and Development Program of China Grant No.2020YFC2201502, grants from NSFC (grant No. 11975019, 11991052, 12047503), Key Research Program of Frontier Sciences, CAS, Grant NO. ZDBS-LY-7009, CAS Project for Young Scientists in Basic Research YSBR-006, the Key Research Program of the Chinese Academy of Sciences (Grant NO. XDPB15).
ZCC is supported by the fellowship of China Postdoctoral Science Foundation No. 2022M710429.
RMS acknowledges support through Australian Research Council Future Fellowship FT190100155.
This work has been carried out by the Parkes Pulsar Timing Array, which is part of the International Pulsar Timing Array. The Parkes radio telescope (``Murriyang'') is part of the Australia Telescope, which is funded by the Commonwealth Government for operation as a National Facility managed by CSIRO.
\end{acknowledgments}
%%%%%%%%%%%%%%%%%%%%%%%%%%%%%%%%%%%%%%%%%%%%%%%%%%%%%%%%%%%%%%%%%%%%%%%%%%%%%%%%
%%%%%%%%%%%%%%%%%%%%%%%%%%%%%%%%%%% references %%%%%%%%%%%%%%%%%%%%%%%%%%%%%%%%%
%\bibliographystyle{apj}
\bibliography{./bibfile-uvdm}
% \bibliographystyle{unsrt} 
% \bibliographystyle{aasjournal}

%%%%%%%%%%%%%%%%%%%%%%%%%%%%%%%%%%%%%%%%%%%%%%%%%%%%%%%%%%%%%%%%%%%%%%
\end{document}